\documentclass[a4paper,aps,prd,preprint,groupedaddress,showpacs]{revtex4}
\usepackage{amssymb}
\usepackage{graphicx}
\usepackage{bm}

\newcommand {\func}[1]{\,\mathrm{#1}\,}

\begin{document}

\title{Deflection of light in time-periodic spherically symmetric gravitational fields}
\author{Vladimir A. Koutvitsky}
\author{Eugene M. Maslov}
\email{zheka@izmiran.ru}
\affiliation{Pushkov Institute of Terrestrial Magnetism, Ionosphere and Radio Wave Propagation (IZMIRAN) 
of the Russian Academy of Sciences,\\ Moscow, Troitsk, Kaluzhskoe Hwy 4, Russian Federation, 108840}
\date{\today}

\begin{abstract}
Using the geodesic method and the perturbative approach, we study the
deflection of light by time-periodic spherically symmetric gravitational
fields. Assuming the weakness of the gravitational field, we derive general
formulas that determine the deflection angle in the leading order
approximation. The formulas are valid for both time-periodic and static
metrics. Using these results, we calculate the deflection angle of a light
ray passing through a spherically symmetric oscillating distribution of a
self-gravitating scalar field with a logarithmic potential. It turned out
that in this case the deflection angle does not depend on time in the
leading order.
\end{abstract}


\maketitle

\section{Introduction}

The deflection of light in gravitational fields of compact objects is one of
the first fundamental predictions of General Relativity. This phenomenon
underlies gravitational lensing, which is routinely observed by modern
astronomical methods \cite{Schn}. According to the basic concepts of General
Relativity, photons in a curved spacetime move along null geodesics $x^{\mu
}=x^{\mu }(\lambda )$ satisfying the equation

\begin{equation}
\frac{d^{2}x^{\mu }}{d\lambda ^{2}}+\Gamma _{\alpha \beta }^{\mu }\frac{%
dx^{\alpha }}{d\lambda }\frac{dx^{\beta }}{d\lambda }=0,  \label{eq1}
\end{equation}%
where $\lambda $ is an affine parameter. Equation (\ref{eq1}) is a system of
ordinary differential equations of the second order. It follows that any
given initial conditions $x^{\mu }$ and $dx^{\mu }/d\lambda $ determine the
unique geodesic and, therefore, all its characteristics, including the
deflection angle when passing near a gravitating mass.

For static spacetimes the geodesic method is well developed. Solving (\ref{eq1}) 
leads to the expression for the deflection angle through definite,
usually elliptic, integrals. These integrals can be calculated in the weak
field limit by expanding the integrands in the inverse powers of the large
distance of closest approach or large impact parameter (see, e.g., \cite{Wein}). %
By this procedure the corresponding expansions for the deflection 
angle in Schwarzschild, Reissner-Nordstrom, Kerr, Kerr-Newman, and many
others spacetimes were obtained \cite{Keet,Wei,Batic,Ghaffarnejad,Renzini,Chiba,Cao} 
(see \cite{Jia} for a recent
review). In the leading order the deflection angle in the equatorial plane
of asymptotically flat static spacetimes is given by

\begin{equation}
\Delta \varphi =\frac{4GM}{b}+O((r_{g}/b)^{2}),  \label{eq2}
\end{equation}%
where $G$ is the gravitational constant, $M$ is the total deflecting mass, $%
b $ is the impact parameter, $r_{g}=2GM$. For the ray passing in the
vicinity of the solar limb this angle makes up about $1.75^{\prime \prime }$.

In addition, we note that a different method for calculating the deflection
angle based on the Gauss-Bonnet theorem has recently been proposed for
static spacetimes \cite{Gibb}. In this approach, the deflection of light
rays is considered as a global topological effect. It was later shown that
for asymptotically flat spacetimes this method is equivalent to the method
of geodesics (see \cite{Li} and references therein).

As for the deflection of light in time-dependent metrics, there are
currently only a few works devoted to this issue. For example, in Ref. \cite%
{Dam}, the light deflection by gravitational waves was studied. It was shown
that the time-dependent part of the deflection of light by a localized
source of gravitational waves appears only in high orders of expansion in $%
1/b$. This result was confirmed in \cite{Kop1}, where a general formalism
was developed for calculating the deflection angle in the field of
gravitational waves. An exhaustive study of the light propagation in the
gravitational field of an ensemble of arbitrarily moving and spinning
point-like masses was carried out in \cite{Kop2, Kop3} using retarded Li\'{e}nard-Wiechert potentials. 
By this technique an explicit formula was obtained there 
for the deflection angle of a light ray as a function of time. 
In Ref. \cite{Piat}, an analysis of the effect of cosmological expansion on
light ray deflection was carried out 
on the basis of the McVittie metric as a geometric
description of a point-like deflecting mass embedded in an expanding
universe. It was found that even in the case of a non-constant Hubble
parameter, the time-dependent cosmological background does not affect the
deflection angle in the leading order.

In this paper, we study the deflection of light by a continuous spherically
symmetric matter distribution that 
performs radial pulsations. Since matter under consideration is assumed to 
be completely transparent, a light ray passing through the bulk of the distribution 
is affected only by the time-dependent gravitational field. As the distribution 
of matter, we take the spherically symmetric configuration of a real nonlinear scalar field,
which can be considered as a simplified model of oscillating halos of dark matter or
oscillating soliton stars formed due to fragmentation and  relaxation processes %
\cite{Seidel1, Seidel2}. 

The nature of dark matter (DM) is still unknown. One of the most cited possibilities 
is that dark matter is composed of ultralight bosonic
particles, axions, with masses in the $10^{-23}\div 10^{-18}$ eV range. 
Due to the huge occupation number, the ensemble of these particles in a coherent 
state can be considered as a classical scalar field, oscillating near the minimum 
of the effective potential. Fragmentation of this scalar condensate can lead to 
the formation of the oscillating localized dark matter objects. 
Oscillations of the scalar field in these objects cause oscillations 
of the gravitational potential, which can be detected by their effect on the motion 
of photons and test bodies. 
Several new methods of probing the DM,  pointed out on detection of scalar field 
oscillations, were proposed lately for the case of DM interacting with the photons 
and barionic matter through gravitation only \cite{Rubakov, Aoki, Sibir, Bosk}. 
In particular, the gravitational field of the oscillating galactic halo composed 
out of such dark matter was shown to cause small periodic  fluctuations in the observed 
timing array of the pulsar located inside the halo. The effect is due to the gravitational 
time delay for the photon passing through the halo \cite{Rubakov}.  
While predicted effect is very small, the authors believe it can be spotted since the 
frequency of the gravitation field oscillations is in the range of pulsar timing array 
observations. An appoach based on the interferometer experiments was proposed in the paper \cite{Aoki} 
for detecting the ultralight axion dark matter wind due to the motion of the Earth through DM.  
Oscillations of  the DM produce gravitational field oscillations, which look like gravitational 
waves for the observer on Earth and would be detected by future laser interferometer experiments. 
In Ref. \cite{Sibir} observations of binary pulsars was proposed as a probe of ultralight axion dark matter. 
Oscillations of DM were shown to perturb orbits of binary pulsars, and the exquisite precision 
of measurements  makes the binary pulsars highly sensitive detector of these effects. 
Also, in the context of ultralight dark matter, where galactic halos produce a time-dependent 
and periodic gravitational potential, motion of stars was considered in spherically symmetric 
but time-dependent backgrounds  \cite{Bosk}. It was  shown that orbital resonances may occur and that spectroscopic 
emission lines from stars in these geometries exhibit characteristic, periodic modulation patterns. 
The obtained results applied to the galactic center show that the motion of S2-like 
stars may carry distinguishable observational imprints of the oscillating DM.

The above motivates us to study deflection of light in time-dependent
gravitational fields. We expect that the deflection angle will be modulated
by the frequency of the gravitational field oscillations. This effect can
result in the corresponding periodical variations of intensity of light from
distant sources and add to lightcurves in microlensing observations.

Note that in the static case, the deflection of light by
localized scalar field configurations was considered by several authors. In
Ref. \cite{Vir}, the gravitational deflection of light by the static
spherically symmetric distribution of a massless scalar field was studied on
the basis of the Janis--Newman--Winicour solution \cite{Jan}. As a result, the
deflection angle was analytically calculated by the geodesic method. In Ref. 
\cite{Dab}, gravitational lensing by a static spherically symmetric boson
star formed by a massive complex scalar field without self-interaction was
considered. The deflection angle was calculated there using numerical
solution of the Einstein-Klein-Gordon system. The gravitational lensing by a
static spherically symmetric halo constructed of the nonlinear complex
scalar field with a $\phi ^{6}$-type self-interaction was studied in \cite{Sch}. %
Using the well--known Emden solution of the corresponding nonlinear
Klein--Gordon equation in flat spacetime, the authors calculated the
deflection angle in the lowest approximation. 

Our paper is organized as follows. In Sec. II, assuming the weakness of the
gravitational field, we use a perturbative approach for calculating the
leading order term of the deflection angle in nonstatic spherically symmetric
spacetimes. In Sec. III, we apply the obtained formulas to
calculate the deflection angle of a light ray in a time-dependent
gravitational field of an oscillating configuration of a real scalar field
with a logarithmic self--interaction. Some remarks concerning the obtained
results are made in Sec. IV.

\section{The geodesic method for nonstatic spherically symmetric spacetimes}

Let us consider a spherically symmetric nonstatic metric of the form%
\begin{equation}
ds^{2}=B(t,r)\,dt^{2}-A(t,r)\,dr^{2}-r^{2}(d\vartheta ^{2}+\sin
^{2}\vartheta \,d\varphi ^{2}).  \label{eq3}
\end{equation}%
For light rays lying in the plane $\vartheta =\pi /2$, equation (\ref{eq1})
reduces to the system%
\begin{equation}
\frac{d^{2}t}{d\lambda ^{2}}+\frac{\dot{B}}{2B}\left( \frac{dt}{d\lambda }%
\right) ^{2}+\frac{B^{\prime }}{B}\frac{dt}{d\lambda }\frac{dr}{d\lambda }+%
\frac{\dot{A}}{2B}\left( \frac{dr}{d\lambda }\right) ^{2}=0,  \label{eq4}
\end{equation}%
\begin{equation}
\frac{d^{2}r}{d\lambda ^{2}}+\frac{B^{\prime }}{2A}\left( \frac{dt}{d\lambda 
}\right) ^{2}+\frac{\dot{A}}{A}\frac{dt}{d\lambda }\frac{dr}{d\lambda }+%
\frac{A^{\prime }}{2A}\left( \frac{dr}{d\lambda }\right) ^{2}-\frac{r}{A}%
\left( \frac{d\varphi }{d\lambda }\right) ^{2}=0,  \label{eq5}
\end{equation}%
\begin{equation}
\frac{d^{2}\varphi }{d\lambda ^{2}}+\frac{2}{r}\frac{dr}{d\lambda }\frac{%
d\varphi }{d\lambda }=0,  \label{eq6}
\end{equation}%
where 
$(\dot{\phantom{.}})=\partial /\partial t$, $\left( ^{\prime
}\right) =\partial /\partial r$. From Eq. (\ref{eq6}) it follows that for a
ray coming from infinity%
\begin{equation}
\frac{d\varphi }{d\lambda }=\frac{b}{r^{2}},  \label{eq7}
\end{equation}%
where $b$ is the impact parameter. Another integral of motion is obtained
from (\ref{eq3}) if we put $ds^{2}=0,$ $\vartheta =\pi /2$ and use Eq. (\ref{eq7}). 
This gives%
\begin{equation}
B\left( \frac{dt}{d\lambda }\right) ^{2}-A\left( \frac{dr}{d\lambda }\right)
^{2}-\frac{b^{2}}{r^{2}}=0.  \label{eq8}
\end{equation}%
Indeed, differentiating Eq. (\ref{eq8}) with respect to $\lambda $ and using
(\ref{eq4}), (\ref{eq7}), we get Eq. (\ref{eq5}).

In addition, note that Eq. (\ref{eq4}) can be rewritten as%
\begin{equation}
\frac{d}{d\lambda }\ln \left( B\frac{dt}{d\lambda }\right) =\frac{\dot{B}}{2B%
}\frac{dt}{d\lambda }-\frac{\dot{A}}{2B}\left( \frac{dr}{d\lambda }\right)
^{2}\left( \frac{dt}{d\lambda }\right) ^{-1}.  \label{eq9}
\end{equation}

Let us assume that the gravitational field is time--periodic (with a certain
period $T_{g}$) and weak everywhere on the light ray, i.e.,%
\begin{equation}
A=1-2\psi +O(\varkappa ^{2}),\quad B=1+2\chi +O(\varkappa ^{2}),
\label{eq11}
\end{equation}%
where $\psi (t,r)$ and $\chi (t,r)$ are time--periodic functions of order $%
\varkappa \ll 1$, and $\varkappa $ is a dimensionless small parameter
proportional to the gravitational constant $G$.

Now suppose that in the $xy$ plane at a distant point $x=x_{0}=nT_{g}$, $y=b$
($n$ is a large integer) at a moment $t_{0}$, a photon is emitted parallel
to the $x$ axis in the direction of the gravitating mass. If the gravitating
mass were absent, the photon would move along the straight line%
\begin{equation}
x=x_{0}+t_{0}-t,\quad y=b  \label{eq12}
\end{equation}%
with the current radial coordinate%
\begin{equation}
r(t)=\sqrt{x^{2}(t)+b^{2}}  \label{eq13}
\end{equation}%
and be registered at the distant point $x=-x_{0}$, $y=b$ at the moment $%
t_{R}=t_{0}+2x_{0}$. On this trajectory we can set $t=\lambda $.

With the gravitating mass, the photon will move along a deflected trajectory
with the current radial coordinate%
\begin{equation}
r(t)=\left( 1+\eta (t)\right) \sqrt{x^{2}(t)+b^{2}},  \label{eq14}
\end{equation}%
where, as before,%
\begin{equation}
x=t_{R}-t-x_{0},  \label{eq15}
\end{equation}%
and $\eta (t)$ is a small function of order $\varkappa $. On this
trajectory, the dependence $t(\lambda )$ is determined by Eq. (\ref{eq9}),
where we set%
\begin{equation}
B\frac{dt}{d\lambda }=1+\zeta (t)  \label{eq16}
\end{equation}%
with a small function $\zeta (t)\sim \varkappa $. Then, with the required
accuracy, from Eq. (\ref{eq9}) we obtain%
\begin{equation}
\frac{d\zeta }{dt}=\dot{\chi}(t,r)+\dot{\psi}(t,r)\left( 1-\frac{b^{2}}{r^{2}%
}\right) ,  \label{eq17}
\end{equation}%
where $r(t)$ is defined in (\ref{eq13}), (\ref{eq15}). Setting $d/dt=-d/dx$
and integrating over the interval $(x,x_{0})$, we can find the dependence $%
\zeta (x)$.

Further, taking into account (\ref{eq16}), we rewrite Eq. (\ref{eq8}) in the
form%
\begin{equation}
1-\frac{A}{B}\left( \frac{dr}{dt}\right) ^{2}=\frac{b^{2}B}{r^{2}\left(
1+\zeta \right) ^{2}},  \label{eq18}
\end{equation}%
where $r(t)$ is now given by (\ref{eq14}), (\ref{eq15}). Differentiation of
Eq. (\ref{eq14}) gives%
\begin{equation}
\left( \frac{dr}{dt}\right) ^{2}=\frac{x^{2}}{x^{2}+b^{2}}\left( 1+2\eta
\right) -2x\frac{d\eta }{dt}+O(\varkappa ^{2}).  \label{eq19}
\end{equation}%
Substituting Eqs. (\ref{eq11}) and (\ref{eq19}) into Eq. (\ref{eq18}), we
arrive at the equation%
\begin{eqnarray}
&&x(x^{2}+b^{2})\frac{d\eta }{dt}-(x^{2}-b^{2})\eta  \nonumber \\
&=&-x^{2}\psi (t,r)-(x^{2}-b^{2})\chi (t,r)-b^{2}\zeta (t),  \label{eq20}
\end{eqnarray}%
where $r(t)$ and $x(t)$ are given by (\ref{eq13}) and (\ref{eq15}). The
solution to this equation is%
\begin{widetext}
\begin{equation}
\eta =\frac{x}{x^{2}+b^{2}}
\left\{ \int \left[ x^{2}\psi(t,r)+(x^{2}-b^{2})\chi (t,r)+b^{2}\zeta (t)\right] \frac{dx}{x^{2}}+const\right\} ,  \label{eq21}
\end{equation}
\end{widetext}
where $dx=-dt$ in accordance with Eq. (\ref{eq15}). The $constant$ in (\ref%
{eq21}) can be found from the condition $\eta (x_{0})=0$, but, as we will
see below, it does not affect the deflection angle when $x_{0}\rightarrow
\infty $.

Now consider Eq. (\ref{eq7}). Using Eqs. (\ref{eq14}) and (\ref{eq16}), in
the first order in $\varkappa $ we find%
\begin{equation}
\frac{d\varphi }{dt}=\frac{b}{x^{2}+b^{2}}\left[ 1+\left( 2\chi -\zeta
-2\eta \right) \right] .  \label{eq22}
\end{equation}%
Setting $d/dt=-d/dx$, we integrate (\ref{eq22}) over the interval $\left(
-x_{0},x_{0}\right) $ and take the limit $x_{0}=nT_{g}\rightarrow \infty $ $%
(n\rightarrow \infty )$. As a result, we obtain $\varphi =\pi +\Delta
\varphi $, where the deflection angle in the leading order is given by%
\begin{equation}
\Delta \varphi =b\int_{-\infty }^{\infty }\frac{2\chi -\zeta -2\eta }{%
x^{2}+b^{2}}\,dx.  \label{eq23}
\end{equation}

Note that this formula is valid not only for time--periodic metrics, but also
for %
the
static ones. Indeed, we can consider the static metric as a limiting
case of time--periodic as $T_{g}\rightarrow \infty $, and therefore we should
put $\zeta =0$ in accordance with Eq. (\ref{eq17}). Consider, for example,
the Schwarzschild metric. Assuming $r_{g}/b=\varkappa \ll 1$, where $%
r_{g}=2GM$ is the gravitational radius, we have%
\begin{equation}
\psi =\chi =-\varkappa \frac{b}{2r}.  \label{eq24}
\end{equation}%
Then formula (\ref{eq21}) gives%
\begin{equation}
\eta =-\varkappa \frac{bx}{x^{2}+b^{2}}\left( \frac{\sqrt{x^{2}+b^{2}}}{2x}+%
\func{arsh}\frac{x}{b}+const\right) .  \label{eq25}
\end{equation}%
Substituting (\ref{eq24}) and (\ref{eq25}) into (\ref{eq23}) and
integrating, we obtain the well-known result (\ref{eq2}).

In the case of a time--periodic metric, the deflection angle will generally
depend on the photon emission time $t_{0}$ or, which is the same, on the
observation time $t_{R}=t_{0}+2x_{0}$. Indeed, in the calculations we
substitute $t=t_{R}-x_{0}-x$ with $x_{0}=nT_{g}$ into the $T_{g}$--periodic
gravitational potentials $\psi (t,r)$ and $\chi (t,r)$. Therefore, the
moment $t_{R}$ determines in which phase of the oscillations of the
gravitational field the photon passed through the matter distribution and,
consequently, at what angle it deflected as a result of this.

In the next section, we calculate the deflection angle of a light ray
passing through the oscillating distribution of a real scalar field with a
logarithmic self--interaction.

\section{Deflection of light by a time-periodic spherically symmetric scalar
field}

As a deflecting matter, we consider the self--gravitating real scalar field
with the potential%
\begin{equation}
U(\phi )=\frac{m^{2}}{2}\phi ^{2}\left( 1-\ln \frac{\phi ^{2}}{\sigma ^{2}}%
\right) ,  \label{eq26}
\end{equation}%
where $\sigma $ is the characteristic magnitude of the field, $m$ is the
mass (in units $\hbar =c=1)$. Such potentials arise in quantum field theory 
\cite{Rosen, Birula}, in inflationary cosmology \cite{Barrow}, and also in
some supersymmetric extensions of the Standard Model \cite{Enq}. They allow
the existence of localized time--periodic field configurations, the pulsons
(oscillons, in another terminology) \cite{Marq, Bog, Mas}. The corresponding
solution of the Einstein--Klein--Gordon system was found in Ref. \cite{Koutv1}
in the weak field approximation. It has the form%
\begin{equation}
\phi (t,r)=\sigma \lbrack a(\theta )+\varkappa Q(\theta ,\rho )+O(\varkappa
^{2})]e^{(3-\rho ^{2})/2},  \label{eq27}
\end{equation}%
\begin{equation}
A(t,r)=\left( 1-\frac{\rho _{g}}{\rho }\right) ^{-1},\quad B(t,r)=\left( 1-%
\frac{\rho _{g}}{\rho }\right) e^{-s},  \label{eq28}
\end{equation}%
where%
\begin{eqnarray}
\rho _{g}(\tau ,\rho )&=&-\varkappa \rho \left[ V_{\max }\left( 1-\frac{\sqrt{%
\pi }\func{erf}\rho }{2\rho }e^{\rho ^{2}}\right) +a^{2}\rho ^{2}\right]
e^{3-\rho ^{2}}\nonumber\\
&+&O(\varkappa ^{2}),\\  \label{eq29}
s(\tau ,\rho )&=&\varkappa (2V_{\max }+a^{2}\ln a^{2}+a^{2}\rho ^{2})e^{3-\rho^{2}}\!\!+\! %
O(\varkappa ^{2}),  \label{eq30}
\end{eqnarray}
$\tau =mt$, $\rho =mr$, $\varkappa =4\pi G\sigma ^{2}\ll 1$ ($G$ is the
gravitational constant). The function $a(\theta (\tau ))$ oscillates in the
range $-a_{\max }\leqslant a(\theta )\leqslant a_{\max }$ in the local
minimum of the potential $V(a)$:%
\begin{equation}
a_{\theta \theta }=-dV/da,  \label{eq31}
\end{equation}%
\begin{equation}
V(a)=(a^{2}/2)\left( 1-\ln a^{2}\right) \leqslant V_{\max }=V(a_{\max }),
\label{eq32}
\end{equation}%
where $\theta _{\tau }=1+\varkappa \Omega +O(\varkappa ^{2})$, and the
constant $\varkappa \Omega $ is the pulson frequency correction due to
gravitational effects. The period of these oscillations is given by%
\begin{equation}
T=4\int_{0}^{1}\left[ (1-\ln a_{\max }^{2})(1-z^{2})+z^{2}\ln z^{2}\right]
^{-1/2}dz.  \label{eq33}
\end{equation}%
The function $Q(\theta ,\rho )$ is a series in Hermite polynomials whose
coefficients are $T$--periodic (in $\theta $) solutions of nonhomogeneous
Hill equations. In Ref. \cite{Koutv1}, initial conditions and the correction 
$\varkappa \Omega $ were found for which such solutions exist. The stability
of these solutions essentially depends on the oscillation amplitude $a_{\max
}$. It turned out that in some intervals of $a_{\max }$ values, solutions
with high accuracy retain their periodicity, making hundreds of
oscillations. Similar quasistability intervals were also found when studying
the effect of external perturbations on the pulson in the absence of gravity 
\cite{Koutv2, Koutv3}. In what follows, we assume that $a_{\max }$ belongs
to one of these intervals.

In addition, we note that potential (\ref{eq26}) also admits localized pulsating
solutions, the nodal pulsons, which, in contrast to the Gaussian--like
nodeless pulsons, have a decaying wave structure at the periphery. It turned
out, however, that such pulsons are highly unstable at any values of the
amplitudes. Both nodal and nodeless pulsons can arise from arbitrary
oscillating initial conditions. For example, as shown in Ref. \cite{Koutv3}, a
homogeneous oscillating scalar condensate (of the Affleck-Dine type) decays
due to parametric instability into an ensemble of pulsons, of which only
nodeless (Gaussian-like) pulsons with amplitudes lying in the intervals of
quasistability survive at the final stage. It is these pulsons that can be
of astrophysical interest as decay products of oscillating dark matter. The
above solution describes such pulsons taking into account the self--gravity
effects. They have the characteristic radius $\sim 1/m$ and pulsate with the
period $[m(1+\varkappa \Omega )]^{-1}T$ (with respect to $t$).
Accordingly, the gravitational field inside the pulson varies with
the period $T_{g}=[2m(1+\varkappa \Omega )]^{-1}T$. Recently, in Ref. \cite%
{Koutv4}, we examined the gravitational frequency shift of a light signal
from the center of this pulson and found its periodic variations. Now we
consider the deflection of a light ray passing through the pulson.

Since the metric found is everywhere regular and has no horizon, we can
rewrite the functions $A(t,r)$ and $B(t,r)$ with the required accuracy in
the form (\ref{eq11}), where%
\begin{equation}
\psi (t,r)=\frac{\varkappa }{2}\left[ V_{\max }\left( 1-\frac{\sqrt{\pi }%
\func{erf}\rho }{2\rho }e^{\rho ^{2}}\right) +a^{2}\rho ^{2}\right]
e^{3-\rho ^{2}},  \label{eq34}
\end{equation}%
\begin{equation}
\chi (t,r)=-\frac{\varkappa }{2}\left[ V_{\max }\left( 1+\frac{\sqrt{\pi }%
\func{erf}\rho }{2\rho }e^{\rho ^{2}}\right) +a^{2}\ln a^{2}\right]
e^{3-\rho ^{2}}.  \label{eq35}
\end{equation}%
Calculating $\dot{\psi}(t,r)$, $\dot{\chi}(t,r)$ and setting

\begin{eqnarray*}
\tau &=&\tau _{R}-\xi _{0}-\xi ,\quad \xi =mx,\quad \xi _{0}=mx_{0},\quad
\tau _{R}=mt_{R}, \\
\beta &=&mb,\quad \quad \qquad \rho ^{2}=\xi ^{2}+\beta ^{2},\quad \quad
\qquad d/d\tau =-d/d\xi ,
\end{eqnarray*}%
from Eq. (\ref{eq17}) we find%
\begin{equation}
\zeta =\frac{\varkappa }{2}e^{3-\beta ^{2}}\int_{\xi }^{\xi _{0}}\left[ 
\frac{d}{d\xi }\left( a^{2}\ln a^{2}\right) -\xi ^{2}\frac{d}{d\xi }a^{2}%
\right] e^{-\xi ^{2}}d\xi ,  \label{eq36}
\end{equation}%
where $\xi _{0}\rightarrow \infty $.

On the other hand, from Eqs.(\ref{eq31}) and (\ref{eq32}) it follows that%
\[
\frac{d^{2}a^{2}}{d\xi ^{2}}=\frac{d^{2}a^{2}}{d\tau ^{2}}=\frac{d^{2}a^{2}}{%
d\theta ^{2}}\theta _{\tau }^{2} 
\]%
\begin{equation}
=4V_{\max }-2a^{2}+4a^{2}\ln a^{2}+O(\varkappa ).  \label{eq36a}
\end{equation}%
We will use this relation to exclude $a^{2}\ln a^{2}$ from calculations.
Thus, integrating in (\ref{eq36}) by parts and using Eq. (\ref{eq36a}), we
obtain%
\begin{equation}
\zeta =-\frac{\varkappa }{4}e^{3-\rho ^{2}}\left( \frac{1}{2}\frac{d^{2}}{%
d\xi ^{2}}+\xi \frac{d}{d\xi }\right) a^{2}+O(\varkappa ^{2}).  \label{eq38}
\end{equation}

Now we substitute $\psi $, $\chi $ and $\zeta $ into Eq. (\ref{eq21}), use
Eq. (\ref{eq36a}) and integrate over $\xi $ by parts. This gives%
\begin{eqnarray}
\eta &=&\frac{\varkappa }{2}e^{3-\beta ^{2}}\Bigg{\{}\sqrt\pi\, V_{\max }\left[ 
\frac{\xi \func{erf}\xi }{\rho ^{2}}-e^{\beta ^{2}}\left( \frac{\xi }{\rho
^{2}}\int_{0}^{\xi }\frac{\func{erf}\rho }{\rho }\,d\xi +\frac{\func{erf}%
\rho }{2\rho }\right) \right] \nonumber\\%
&\phantom{=}&-\frac{1}{2}e^{-\xi ^{2}}\left( a^{2}+\frac{\xi }{2\rho ^{2}}\frac{da^{2}}
{d\xi }\right) +const\,\frac{\xi }{\rho ^{2}}\Bigg{\}}+O(\varkappa ^{2}),
\label{eq39}
\end{eqnarray}%
where the identity%
\begin{equation}
\beta ^{2}\int \frac{\func{erf}\rho }{\rho }\frac{d\xi }{\xi ^{2}}=e^{-\beta
^{2}}\frac{2}{\sqrt{\pi }}\int e^{-\xi ^{2}}d\xi -\frac{\rho \func{erf}\rho 
}{\xi }  \label{eq39a}
\end{equation}%
was used and the indefinite integrals of regular functions were replaced by
the definite integrals over the interval $(0,\xi )$ plus constants.

We turn now to Eq. (\ref{eq23}). Using Eqs. (\ref{eq35}), (\ref{eq38}), (\ref%
{eq39}) and taking into account (\ref{eq36a}), we find
\begin{eqnarray}
2\chi -\zeta -2\eta &=&\varkappa\, e^{3-\beta ^{2}}\Bigg{\{}\sqrt{\pi}\, V_{\max }\frac{%
\xi }{\rho ^{2}}\left[ e^{\beta ^{2}}\!\!\int_{0}^{\xi }\frac{\func{erf}\rho }{%
\rho }\,d\xi -\func{erf}\xi \right] \phantom{-} \nonumber\\
&\phantom{=}&-\frac{1}{8}\frac{d^{2}a^{2}}{d\xi ^{2}}e^{-\xi ^{2}}+\frac{1}{4}\left( 1+%
\frac{1}{\rho ^{2}}\right) \frac{da^{2}}{d\xi }\xi e^{-\xi ^{2}}+const\,%
\frac{\xi }{\rho ^{2}}\Bigg{\}}+O(\varkappa ^{2}).  \label{eq40}
\end{eqnarray}
It is remarkable that in this expression the last three terms, two of which
contain derivatives of the oscillating function $a^{2}$, do not contribute
to the integral in Eq. (\ref{eq23}). Integration of the remaining terms gives%
\[
\int_{-\infty }^{\infty }\frac{\xi }{\rho ^{4}}\left( \int_{0}^{\xi }\frac{%
\func{erf}\rho }{\rho }\,d\xi \right) \,d\xi =\frac{1}{\beta ^{2}}\left(
1-e^{-\beta ^{2}}\right) 
\]%
\begin{equation}
+\frac{2}{\sqrt{\pi }}e^{-\beta ^{2}}\int_{0}^{\infty }\frac{e^{-\xi ^{2}}}{%
\xi ^{2}+\beta ^{2}}\,d\xi ,  \label{eq40a}
\end{equation}%
\begin{equation}
\int_{-\infty }^{\infty }\frac{\xi \func{erf}\xi }{\rho ^{4}}\,d\xi =\frac{2%
}{\sqrt{\pi }}\int_{0}^{\infty }\frac{e^{-\xi ^{2}}}{\xi ^{2}+\beta ^{2}}%
\,d\xi .  \label{eq40b}
\end{equation}%
As a result, we arrive at a simple formula for the deflection angle:%
\begin{eqnarray}
\Delta \varphi &=&\varkappa \frac{e^{3}\sqrt{\pi }V_{\max }}{\beta }\left(
1-e^{-\beta ^{2}}\right) +O\left( \varkappa ^{2}\right)  \nonumber \\
&=&\frac{4GM}{b}\left( 1-e^{-m^{2}b^{2}}\right) +O\left( \varkappa
^{2}\right) ,  \label{eq41}
\end{eqnarray}%
where $M$ is the pulson mass,%
\begin{equation}
M=\left( e\sqrt{\pi }\right) ^{3}\sigma ^{2}m^{-1}V_{\max }\left(
1+O(\varkappa )\right) .  \label{eq42}
\end{equation}%
This formula is valid for any values of the impact parameter, $0\leqslant
b<\infty $. In particular, $\Delta \varphi =0$ for $b=0$, which is quite
natural. The maximum of $\Delta \varphi $ is achieved at $mb=1.1209$.

\section{Concluding remarks}

In this paper we have considered deflection of light rays in time--dependent
spherically symmetric gravitational fields. Working in the weak field
approximation, we have obtained general formulas that determine the
deflection angle of a light ray coming from a distant source, passing
through the oscillating matter distribution, and observed by a distant
observer (see Eqs. (\ref{eq17}), (\ref{eq21}) and (\ref{eq23})). These
formulas are valid not only for time-periodic metrics, but also for static
ones.

We used these results to calculate the deflection angle of light passing
through a pulsating lump of the self--gravitating scalar field with the
logarithmic self--interaction. It is interesting that in this case the
deflection angle turned out to be time--independent in the leading order,
despite the oscillations of the scalar field (see Eq. (\ref{eq41})). We
believe that this is a specific feature of the logarithmic potential.
Indeed, suppose for a moment that the function $a(\theta )$ in Eqs. (\ref%
{eq34}) and (\ref{eq35}) satisfies Eq. (\ref{eq31}) with some potential $V(a)
$ different from (\ref{eq32}). This means that the space--time geometry is
given a priori. Then the additional term 
\[
\varkappa \beta e^{3-\beta ^{2}}\int_{-\infty }^{\infty }\left( V(a)+\frac{a%
}{2}\frac{dV}{da}-\frac{a^{2}}{2}+a^{2}\ln a^{2}\right) e^{-\xi ^{2}}d\xi 
\]%
appears on the right hand side of formula (\ref{eq41}), which obviously
depends on time of observation $t_{R}$. This term vanishes only with the
logarithmic potential (\ref{eq32}).

In general case the deflection angle $\Delta \varphi $ (22) is of course
time--dependent. In observations it would manifest itself in additional
variations of intensity of images when lensing the distant sources. The
intensity of the image is described by the gain factor \cite{Ohan1}. In weak
lensing it can be estimated as%
\[
(gain)\approx \frac{2b}{D^{2}}\left\vert \frac{d(\Delta \varphi )^{2}}{db}%
\right\vert ^{-1}, 
\]%
where $D\approx D_{o}D_{s}(D_{o}+D_{s})^{-1}$, $D_{o}$ and $D_{s}$ are the
distances from the observer to the lens and from the lens to the source,
respectively. The variations of the time--dependent deflection angle $\Delta
\varphi $ have the period $T_{g}$ of the lens gravitational field
oscillations. If this period is much less than the characteristic time-scale
of the change of the relative positions of the source, the lens, and the
observer, the intensity of the image will be also time-periodic with the
same period.

In the case of lensing by the pulsating scalar object with the logarithmic
self--interaction, we find%
\[
(gain)\approx \frac{\left( 4GMm^{2}D\right) ^{-2}\beta ^{4}}{\left(
1-e^{-\beta ^{2}}\right) \left\vert 1-\left( 1+2\beta ^{2}\right) e^{-\beta
^{2}}\right\vert }. 
\]%
Note that this gain factor becomes infinite at $\beta =1.1209$ where $\Delta
\varphi $ (43) has the maximum. In fact, in such cases, the intensity of the
image becomes large but not infinite due to diffraction effects \cite{Ohan2}.

\section*{Acknowledgment}
We would like to thank the referee for criticism and useful comments, 
as well for drawing our attention to paper \cite{Ohan1}.

\end{document}